\documentclass[12pt]{article}

\usepackage{amssymb,amsmath,mathrsfs}
\newtheorem{theorem}{Theorem}
\newtheorem{remark}{Remark}
\newtheorem{lemma}{Lemma}
\newtheorem{corollary}{Corollary}

\begin{document}
\begin{center}
\sc\sl Andrey Novikov\footnote{Departamento de Matem\'aticas, Universidad Aut\'onoma Metropolitana - Iztapalapa,
 San Rafael Atlixco 186, col. Vicentina, C.P. 09340, M\'exico D.F., M\'exico; email: {\tt an@xanum.uam.mx}}, Petr Novikov\footnote{Kazan State University, Department of Mathematical Statistics,  Kremlevskaya 18, Kazan, Russian Federation; email: {\tt pnovi@mail.ru}}
\end{center}
\begin{center}
{
LOCALLY MOST POWERFUL SEQUENTIAL TESTS OF A SIMPLE HYPOTHESIS VS. ONE-SIDED ALTERNATIVES FOR INDEPENDENT OBSERVATIONS}
\end{center}\vspace{7mm}

\hspace{2mm}\begin{minipage}{120mm} \small \hspace{3mm} Let $X_1,X_2,\dots, X_n, \dots$ be a stochastic process with independent values whose distribution $P_\theta$ depends on an unknown parameter $\theta$, $\theta\in\Theta$, where $\Theta$
is an open subset of the real line. The problem of testing $H_0:$ $\theta=\theta_0$ vs. a composite alternative
$H_1:$ $\theta>\theta_0$ is considered, where $\theta_0\in\Theta$ is a fixed value of the parameter.
The main objective of this work is the characterization of the structure of the locally most powerful (in the sense of Berk \cite{berk}) sequential tests in this problem.
\end{minipage}\vspace{5mm}\\

\hspace{2mm}\begin{minipage}{120mm} \small \hspace{3mm} {\em Keywords}: sequential analysis, hypothesis testing, one-sided alternative, sequential test, locally most powerful test, optimal sequential decision.
\end{minipage}\vspace{5mm}

{\bf  1. Introduction.}
Let $X_1,X_2,\dots, X_n, \dots$ be a stochastic process with independent values whose distribution, $P_\theta$, depends on an unknown parameter $\theta$, $\theta\in\Theta$, where $\Theta$
is an open subset of the real line. The problem of testing $H_0:$ $\theta=\theta_0$ vs. a composite hypothesis
$H_1:$ $\theta>\theta_0$ is considered, where $\theta_0\in\Theta$ is some fixed value of the parameter.
 The main objective of this work is to characterize the structure of the locally most powerful (in the sense of Berk \cite{berk}) sequential tests in this problem.

We follow \cite{novikov09c} in definitions and notation related to sequential hypothesis testing problems
(see also \cite{wald50}, \cite{ferguson}, \cite{degroot}, \cite{schmitz}, \cite{ghosh}, \cite{novikov09b}, among many others).

In particular, we say that
 $(\psi,\phi)$ is a sequential hypothesis test  with a (randomized) stopping rule $\psi$ and a (randomized) decision rule $\phi$ if
$$\psi=\left(\psi_1,\psi_2, \dots ,\psi_{n},\dots\right)\quad\mbox{and}\quad \phi=\left(\phi_1,\phi_2, \dots ,\phi_n,\dots\right),$$
where the functions
$$\psi_n=\psi_n(x_1, x_2,\dots, x_n)
\quad\mbox{and}\quad \phi_n=\phi_n(x_1, x_2,\dots, x_n)$$
are measurable and take values in $[0,1]$, for all $n=1,2,\dots$.

At any stage $n=1,2,\dots$, after some data $(x_1,\dots,x_n)$ are observed,
 the value of $\psi_n(x_1,\dots,x_n)$ is understood as the conditional probability
{\em to stop and proceed to decision-making} given that the experiment came to stage $n$ and that the observations obtained up to this stage were $(x_1, x_2, \dots,
x_n).$ the rules $\psi_1,\psi_2,\dots$
are applied sequentially until the experiment  eventually stops.

After the experiment stops at some stage $n\geq 1$
the decision rule  $\phi_n$ is used to make a decision.
The value $\phi_n(x_1,\dots, x_n)$ is understood as the conditional probability
{\em to reject} the null-hypothesis $H_0$ given the observations $(x_1,\dots, x_n)$.

According to the above procedure,
any stopping rule $\psi$ generates a random variable
$\tau_\psi$ ({\em stopping time}), whose distribution is given by
\begin{equation}\label{1.1}
P_\theta(\tau_\psi=n)=E_\theta(1-\psi_1)(1-\psi_2)\dots (1-\psi_{n-1})\psi_n,\quad n=1,2,\dots.
\end{equation}

Here and throughout the article $E_\theta(\cdot)$
denotes the mathematical expectation with respect to the distribution $P_\theta$ of the process $X_1, X_2, \dots$.

In (\ref{1.1}) we suppose that  $\psi_n=\psi_n(X_1,X_2,\dots,X_n)$,
unlike its previous definition as
$\psi_n=\psi_n(x_1,x_2,\dots, x_n)$.
We use this ``duality'' for interpreting any function of observations $F_n$  making use of the following rule which makes its interpretation non-ambiguous.
  If  $F_n$ is any function of observations
($F_n=F_n(x_1,\dots,x_n)$ or $F_n=F_n(X_1,\dots,X_n)$), and its arguments are omitted, then: \begin{itemize}
\item if $F_n$ is under the probability or the expectation sign, then it stands for  $F_n(X_1,\dots,$ $ X_n)$,
\item
otherwise $F_n$ means $F_n(x_1,\dots, x_n)$.
\end{itemize}

As a characteristic of the duration of the sequential experiment {\em the average sample number} is used:
\begin{equation}\label{1.2}
{\mathscr N}_\theta(\psi)=E_\theta\tau_\psi=\begin{cases}\sum_{n=1}^\infty n P_\theta(\tau_\psi=n),\; \mbox{if}\; P_\theta(\tau_\psi<\infty)=1,\cr
\infty,\; \mbox{ otherwise.} \end{cases}
\end{equation}

For a sequential test  $(\psi,\phi)$ let as define {\em the power function}
in $\theta$ as
\begin{equation}\label{1.3}
\beta_\theta(\psi,\phi)=P_{\theta}(\mbox{reject}\; H_0)=\sum_{n=1}^\infty E_\theta(1-\psi_1)
\dots(1-\psi_{n-1})\psi_n\phi_{n}.
\end{equation}
{\em The first type error probability} of the test $(\psi,\phi)$ is defined as
\begin{equation*}
\alpha(\psi,\phi)=\beta_{\theta_0}(\psi,\phi).
\end{equation*}

The main objective of this work is characterization of the tests which maximize the derivative, at $\theta=\theta_0$, of the power function
$\dot\beta_{\theta_0}(\psi,\phi),$
in the class of all such sequential tests  $(\psi,\phi)$, that
\begin{equation}\label{1.4}
\alpha(\psi,\phi)\leq \alpha,
\end{equation}
and
\begin{equation}\label{1.5}
\mathscr N_{\theta_0}(\psi)\leq \mathscr N,
\end{equation}
where $\alpha\in[0,1)$ and $\mathscr N\geq 1$ are some restrictions.
If such test exists, it is called {\em locally most powerful}
(see \cite{berk}, \cite{roters}).
We use in this article a rather general method initially developed for testing of two  simple hypotheses
(see \cite{novikov09b}), then extended to multiple hypothesis testing (see \cite{novikov09a}),  to general statistical problem with Bayes decisions (see \cite{NovikovBayes}, \cite{NovikovTroyes}) and finally
to the problems of locally most powerful tests (see \cite{novikov09c}), all the problems being for the discrete-time stochastic processes.\vspace{5mm}

{\bf 2. Assumptions and notation}.
Let us suppose that
 $X_i$ has a ``density function'' $f_{\theta,i}$
(Radon-Nikodym derivative of its distribution) with respect to some $\sigma$-finite measure $\mu$ on the space of ``values'' of $X_i$, $i=1,2,3,\dots$.

Due to the independence of the observations, for each
 $n=1,2,3,\dots$ the ``vector''
$(X_1,X_2,\dots X_n)$ of the first $n$ observations has a ``joint density''
$$f_\theta^n(x_1,\dots,x_n)=\prod_{i=1}^n f_{\theta,i}(x_i)$$
with respect to the product-measure
$$\begin{array}{ccc}\mu^n&=&\underbrace{\mu\otimes \mu\otimes \dots\otimes\mu}.\\
&&\small n\quad \mbox{times}\end{array}$$

We will assume (when needed) that the following conditions are fulfilled.

Let
\begin{equation}\label{1.5a}
I_j(\theta_0,\theta_1)=E_{\theta_0}\ln\frac{f_{\theta_0,j}(X_j)}{f_{\theta_1,j}(X_j)}
\end{equation}
be the Kullback-Leibler information for $X_j$ for distinguishing between $\theta=\theta_0$ and $\theta=\theta_1$, $j=1,2,\dots$.\vspace{2mm}

{\sc Assumption 1.} {\sl There exist $\delta>0$ and $0<\gamma_1<\infty$ such that
\begin{equation}\label{1.6}
    I_j(\theta_0,\theta)/(\theta-\theta_0)^2\leq \gamma_1
\end{equation}
for all $j=1,2,\dots$ and for all $|\theta-\theta_0|\leq \delta$.}\vspace{2mm}

For independent and identically distributed (i.i.d.) observations Assumption 1 coincides with Assumption 1 in \cite{berk}.\vspace{2mm}

{\sc Assumption 2.} {\sl For every $j\geq 1$ there exists an integrable (with respect to  $\mu$) function $\dot f_{\theta_0,j}$, such that
$$\int\left|f_{\theta,j}-f_{\theta_0,j}-(\theta-\theta_0)\dot f_{\theta_0,j}\right|d\mu=o((\theta-\theta_0))$$
as $\theta\to \theta_0$.}\vspace{2mm}

In essence, Assumption 2 is a  condition of Frech\'et differentiability of the marginal densities
in the space  $L_1(\mu)$  of integrable with respect to  $\mu$ functions (see similar conditions in \cite{novikov06} and in \cite{mueller-funk}).

It is not difficult to see that Assumption 2 guarantees that the power function of any test based on fixed number of observations is differentiable, and that its derivative can be calculated by differentiating under the integral sign. In this way, for i.i.d observations Assumption 2 entails the validity of Assumption 3 in \cite{berk}.\vspace{2mm}

{\sc Assumption 3.} {\sl There exists $0<\gamma_2<\infty$ such that
$$
E_{\theta_0}\left|\frac{\dot f_{\theta_0,j}(X_j)}{f_{\theta_0,j}(X_j)}\right|\leq \gamma_2
$$
for all $j=1,2,\dots$.}\vspace{2mm}

(Here, and throughount the article, we assume that the mathematical expectation with respect to any ``density function''  $f(x)$:
$$Eg(X)=\int g(x)f(x)\,d\mu(x),$$
is understood as
$Eg(X)=\displaystyle \int g(x)f(x)I_{\{f(x)\neq 0\}}\,d\mu(x),$
so we do not need to care about the definition of $g(x)$ on $\{f(x)=0\}$.)

Assumption 3 is weaker than Assumption 4 in
 \cite{berk} for i.i.d. observations, where the finiteness of the Fisher information is required. In particular, if the Fisher information
\begin{equation}\label{2.3a}I_j(\theta_0)=E_{\theta_0}\left({\dot f_{\theta_0,j}(X_j)\over f_{\theta_0,j}(X_j)} \right)^2\leq \gamma_2^2\end{equation} for all $j=1,2,\dots$, then from the H\"older inequality it follows that Assumption 3 is fulfilled. In turn, (\ref{2.3a}) is closely related to Assumption 1, because under very general  conditions of regularity of the statistical experiment
$$ I_j(\theta,\theta+h)\sim I_j(\theta)h^2/2,\; h\to 0. $$

In the case of i.i.d. observations Assumption 3 follows from  Assumption 2, which guarantees the existence of the finite expectation $E_{\theta_0}|\dot f_{\theta_0,j}(X_j)/f_{\theta_0,j}(X_j)|$.

Because the expression of type $(1-\psi_1)\dots(1-\psi_{n-1})\psi_n$ will be needed frequently (see, e.g., (\ref{1.1}), (\ref{1.3})), let us introduce a notation for it:
\begin{equation}\label{1.10}s_n^\psi= (1-\psi_1)\dots(1-\psi_{n-1})\psi_n,\;n=1,2,\dots.\end{equation}
Let also
\begin{equation}\label{1.11}
t_n^\psi=(1-\psi_1)\dots(1-\psi_{n-1}),\; n=1,2,\dots
\end{equation}
($s_1^\psi\equiv \psi_1$ and $t_1^\psi\equiv 1$ by definition).

Let, finally,
$$ S_n^\psi=\{(x_1,\dots,x_n):s_n^\psi(x_1,\dots,x_n)>0\} $$
and
$$ T_n^\psi=\{(x_1,\dots,x_n):t_n^\psi(x_1,\dots,x_n)>0\}. $$

{\bf 3. Differentiability of the power function and informational in-equalities for test characteristics.}
In this section we prove the existence of the derivative of the power function of any test with a finite, under the null-hypothesis, average sample number, and establish information-type inequalities relating that derivative to other characteristics of the test: the average sample number and the type-I error probability.

Let us define the Kullback-Leibler information containing in the observations of the process
$X_1,X_2,\dots, X_n,\dots$ up to a random stopping time defined by the rule  $\psi$, as
\begin{equation}\label{2.4}
I(\theta_0,\theta;\psi)=\sum_{n=1}^\infty E_{\theta_0}s_n^\psi\left(\sum_{j=1}^n\ln\frac{f_{\theta_0,j}}{f_{\theta,j}}\right)
\end{equation}
(note that the random process of observations $X_1,X_2,\dots$ participates in (\ref{2.4}) implicitly, through $s_n^\psi=s_n^\psi(X_1,\dots,X_n)$ and $f_{\theta,j}=f_{\theta,j}(X_j)$,
and so does it in the definition of the information in one observation in (\ref{1.5a})).

The next two lemmas will be useful for estimations related to the Kullback-Leibler information.

The first one is in essence a variant of the Jensen inequality adapted to sequential experiments.
\begin{lemma}\label{l1} Let $G:[0,\infty)\mapsto \mathbb R\cup \{\infty\}$ be any convex function, and let
 $a_n=a_n(x_1,\dots,x_n)$, $b_n=b_n(x_1,\dots,x_n)$, $n=1,2,\dots$, be any two sequences of non-negative measurable functions. Then, if $$0<\sum_{n=1}^\infty E_{\theta_0}s_n^\psi a_n<\infty,$$
 then
 \begin{equation}\label{1.12}
    \frac{\sum_{n=1}^\infty E_{\theta_0}s_n^\psi a_n G(b_n)}{\sum_{n=1}^\infty E_{\theta_0}s_n^\psi a_n}\geq
    G\left(\frac{\sum_{n=1}^\infty E_{\theta_0}s_n^\psi a_n b_n}{\sum_{n=1}^\infty E_{\theta_0}s_n^\psi a_n}\right).
 \end{equation}

\end{lemma}
In particular, applying Lemma \ref{l1} to $G(x)=-\ln(x)$, $a_n\equiv 1$, $b_n=f_\theta^n/f_{\theta_0}^n$, and supposing that $P_{\theta_0}(\tau_\psi<\infty)=\sum_{n=1}^\infty E_{\theta_0}s_n^\psi=1$, we get that
\begin{equation}\label{1.13}
    I(\theta_0,\theta;\psi)\geq -\ln\left({\sum_{n=1}^\infty E_{\theta}s_n^\psi}\right)\geq 0.
\end{equation}

Let now $(\psi,\phi)$ be any sequential test with  $P_{\theta_0}(\tau_\psi<\infty)=1$. Let us suppose that $0<\beta_{\theta_0}(\psi,\phi)<1$. Then
\begin{eqnarray}\nonumber I(\theta_0,\theta;\psi)&=&\beta_{\theta_0}(\psi,\phi)\frac{\sum_{n=1}^\infty E_{\theta_0}s_n^\psi \phi_n (-\ln(b_n))}{\beta_{\theta_0}(\psi,\phi)}\\\label{1.15}
 &+&(1-\beta_{\theta_0}(\psi,\phi))\frac{\sum_{n=1}^\infty E_{\theta_0}s_n^\psi(1- \phi_n) (-\ln(b_n))}{1-\beta_{\theta_0}(\psi,\phi)},
 \end{eqnarray}
  where $b_n=f_\theta^n/f_{\theta_0}^n$. Because of this, applying Lemma \ref{l1} to both fractions on the right-hand side of (\ref{1.15}) we get
 \begin{eqnarray*}
    I(\theta_0,\theta;\psi) &\geq& -\beta_{\theta_0}(\psi,\phi)\ln\frac{\sum_{n=1}^\infty E_{\theta_0}s_n^\psi \phi_n b_n}{\beta_{\theta_0}(\psi,\phi)} \\
    &&- (1-\beta_{\theta_0}(\psi,\phi))\ln\frac{\sum_{n=1}^\infty E_{\theta_0}s_n^\psi(1- \phi_n) b_n}{1-\beta_{\theta_0}(\psi,\phi)} \\
    &\geq& -\beta_{\theta_0}(\psi,\phi)\ln\frac{\sum_{n=1}^\infty E_{\theta}s_n^\psi \phi_n}{\beta_{\theta_0}(\psi,\phi)} \\
    & & -(1-\beta_{\theta_0}(\psi,\phi))\ln\frac{\sum_{n=1}^\infty E_{\theta}s_n^\psi(1- \phi_n)}{1-\beta_{\theta_0}(\psi,\phi)} \\
    &\geq& -\beta_{\theta_0}(\psi,\phi)\ln\frac{\beta_\theta(\psi,\phi)} {\beta_{\theta_0}(\psi,\phi)}
     -(1-\beta_{\theta_0}(\psi,\phi))\ln\frac{1-\beta_\theta(\psi,\phi)}{1-\beta_{\theta_0}(\psi,\phi)},
   \end{eqnarray*}
   that is
   \begin{equation}\label{1.16}
     I(\theta_0,\theta;\psi) \geq  \beta_{\theta_0}(\psi,\phi)\ln\frac {\beta_{\theta_0}(\psi,\phi)}{\beta_\theta(\psi,\phi)}
     +(1-\beta_{\theta_0}(\psi,\phi))\ln\frac{1-\beta_{\theta_0}(\psi,\phi)}{1-\beta_\theta(\psi,\phi)}
   \end{equation}
   (more general information-type inequalities  can be found in \cite{volodin}, see, for example, Lemma 5.1 therein).

In the same way we deduce that if
 $\beta_{\theta_0}(\psi,\phi)=0$, then
\begin{equation}\label{1.17}
    I(\theta_0,\theta;\psi)\geq -\ln({1-\beta_\theta(\psi,\phi)}),
\end{equation}
and if  $\beta_{\theta_0}(\psi,\phi)=1$, then
\begin{equation}\label{1.18}
    I(\theta_0,\theta;\psi)\geq -\ln{\beta_\theta(\psi,\phi)},
\end{equation}

The next lemma (Wald's identity for non-identically distributed summands) is useful, in particular, for estimation of the information on the left-hand side of  (\ref{1.16}).
\begin{lemma}\label{l4} Let $Y_j=Y_j(X_j)$ be non-negative measurable functions of observations $X_j$ such that $E_\theta Y_j<\infty$, $j=1,2,\dots$. Then for any stopping rule $\psi$ с $P_\theta(\tau_\psi<\infty)=1$
\begin{equation}\label{2.6}
    \sum_{n=1}^\infty E_{\theta}s_n^\psi\left(\sum_{j=1}^n  Y_j\right)=\sum_{j=1}^\infty E_\theta Y_jP_{\theta}(\tau_\psi\geq j).
\end{equation}

\end{lemma}

{ P r o o f.}
Let, for brevity,  $E(\cdot)$ and $P(\cdot)$ denote  $E_{\theta}(\cdot)$  and $P_\theta(\cdot)$, respectively, throughout the proof.

Let us suppose that the left-hand side of  (\ref{2.6}) is finite. Then
$$
\sum_{n=1}^\infty E s_n^\psi\left(\sum_{j=1}^nY_j\right)
=\sum_{n=1}^\infty \sum_{j=1}^n E s_n^\psi Y_j=\sum_{j=1}^\infty \sum_{n= j}^\infty E s_n^\psi Y_j
$$
(changing the order of summation is possible due to the finiteness of the source series). It is not difficult to see that  under the conditions of the Lemma
$$
\sum_{n= j}^\infty E s_n^\psi Y_j=E t_j^\psi Y_j.
$$
By virtue of the independence of $t_j$ (see (\ref{1.11})) and  $Y_j$ we have
$$
 E t_j^\psi Y_j= E t_j^\psi E Y_j=E Y_j P(\tau_\psi\geq j),$$
so that
\begin{equation}\label{2.7a}
\sum_{n=1}^\infty E s_n^\psi\left(\sum_{j=1}^nY_j\right)=\sum_{j=1}^\infty E Y_jP(\tau_\psi\geq j).
\end{equation}
Inverting these reasonings, under the supposition that the right-hand side of (\ref{2.6}) is finite,
we see that the equality in
(\ref{2.6}) holds as well.
$\Box$
\begin{corollary}\label{c0} Suppose that $I_j(\theta_0,\theta)<\gamma<\infty$ for all $j=1,2,\dots$. Then for any stopping rule $\psi$ such that $E_{\theta_0}\tau_\psi<\infty$
\begin{equation}\label{2.6a}
    I(\theta_0,\theta;\psi)=\sum_{j=1}^\infty I_j(\theta_0,\theta)P_{\theta_0}(\tau_\psi\geq j).
\end{equation}
\end{corollary}

{ P r o o f.}
Let $Y_j=\ln f_{\theta_0,j}/f_{\theta,j}$, $Y_j^+=\max\{0,Y_j\}$, $Y_j^-=\max\{0,-Y_i\}$.
Since
$$
E_{\theta_0}Y_j^-=E_{\theta_0}\max\left\{0,\ln\frac{f_{\theta,j}}{f_{\theta_0,j}}\right\}\leq
E_{\theta_0}\max\left\{0,\frac{f_{\theta,j}}{f_{\theta_0,j}}-1\right\}$$$$\leq \int |f_{\theta,j}-f_{\theta_0,j}|d\mu\leq 2,
$$
from Lemma \ref{l4}  we obtain
\begin{equation}\label{2.80}
\sum_{n=1}^\infty E_{\theta_0} s_n^\psi\left(\sum_{j=1}^nY_j^-\right)=\sum_{j=1}^\infty E_{\theta_0} (Y_j^-)P_{\theta_0}(\tau_\psi\geq j),
\end{equation}
where the right-hand side of  (\ref{2.80}) is finite, because $\sum_{j=1}^\infty P_{\theta_0}(\tau_\psi\geq j)=E_{\theta_0}\tau_\psi$.
  Now from the condition $I_j(\theta_0,\theta)<\gamma$, $j\geq 1$, it follows that    $E_{\theta_0}Y_j^+<\gamma+2$, $j\geq 1$, therefore from Lemma \ref{l4} we get
\begin{equation}\label{2.70a}
\sum_{n=1}^\infty E_{\theta_0} s_n^\psi\left(\sum_{j=1}^nY_j^+\right)=\sum_{j=1}^\infty E_{\theta_0} (Y_j^+)P_{\theta_0}(\tau_\psi\geq j),
\end{equation}
and the right-hand side of (\ref{2.70a}) is also finite.

Subtracting both sides of
(\ref{2.80}) from the corresponding sides of (\ref{2.70a}) and then applying the subtraction in the summands,  we get (\ref{2.6a}).
$\Box$

Since $\sum_{j=1}^\infty P(\tau_\psi\geq j)=E \tau_\psi$, from Lemma \ref{l4} it follows that under Assumption 1 that
\begin{equation}\label{2.9}
     I(\theta_0,\theta;\psi)\leq \gamma_1 (\theta-\theta_0)^2E \tau_\psi,
\end{equation}
if $|\theta-\theta_0|\leq \delta$.

The following theorem is a consequence of the informational inequality
 (\ref{1.17}) and it is interesting by itself, because gives some bounds for the characteris-tics (the average sample number, the type-I error probability and the derivative of the power function) of  {\em any} sequential hypothesis test.
\begin{theorem}\label{t1} Suppose that Assumption 1 is fulfilled.
Then for any sequential test
 $(\psi,\phi)$ such that  $E_{\theta_0}\tau_\psi<\infty$
 and the derivative $\dot\beta_{\theta_0}(\psi,\phi)$ of the power function  $\beta_\theta(\psi,\phi)$ at $\theta=\theta_0$
 exists, it holds
\begin{equation}\label{2.6aa}
    (\dot \beta_{\theta_0}(\psi,\phi))^2\leq 2\gamma_1 \beta_{\theta_0}(\psi,\phi)(1-\beta_{\theta_0}(\psi,\phi))E_{\theta_0}\tau_\psi.
\end{equation}

\end{theorem}

{ P r o o f.} Because, throughout this proof, the sequential test $(\psi,\phi)$ remains fixed, let us simply denote $\beta_h=\beta_{\theta_0+h}(\psi,\phi)$ for any $h$ and   $\dot\beta_0=(\beta_{\theta}(\psi,\phi))_\theta^\prime|_{\theta=\theta_0}$, supposing that  for $(\psi,\phi)$ the conditions of Theorem \ref{t1} are satisfied. Analogously, let us  simply write $E(\cdot)$  instead of $E_{\theta_0}(\cdot)$.

Let us deduce now from (\ref{2.9}) that $(\dot \beta_0)^2\leq 2\gamma_1 \beta_0(1-\beta_0)E \tau_\psi$, i.e. (\ref{2.6aa}).

Suppose first that $0<\beta_0<1$. Denote
\begin{equation}\label{2.9a}
w(x)=\beta_0\ln\frac{\beta_0}{x}+(1-\beta_0)\ln\frac{1-\beta_0}{1-x},
\end{equation}
where $x\in[0,1]$ (see the right-hand side of the inequality (\ref{1.16})).
From (\ref{1.16}) and (\ref{2.9}) it follows that
\begin{equation}\label{1.19}
   0\leq w(\beta_h)\leq \gamma_1h^2E \tau_\psi,
\end{equation}
so it is obvious, first of all, that $\beta_h\to \beta_0$, $h\to 0$.

Let $\Delta_h\beta=\beta_h-\beta_0$. Then by the Taylor formula for $\ln(1+x)$
$$
w(\beta_h)=-\beta_0 \ln(1+\Delta_h\beta/\beta_0)-(1-\beta_0)\ln(1-\Delta_h\beta/(1-\beta_0))
$$
$$
=(\Delta_h\beta)^2/(2\beta_0)+(\Delta_h\beta)^2/(2(1-\beta_0))+o((\Delta_h\beta)^2)
$$
$$
=(\Delta_h\beta)^2/(2\beta_0(1-\beta_0))+o((\Delta_h\beta)^2),\quad h\to 0,
$$
from which by virtue of  (\ref{1.19}) it follows that
$$ (\Delta_h\beta/h)^2/(2\beta_0(1-\beta_0))+o((\Delta_h\beta/h)^2)\leq \gamma_1 E \tau_\psi,\quad h\to 0,$$
that is, $(\dot\beta_0)^2/(2\beta_0(1-\beta_0))\leq \gamma_1 E \tau_\psi$, which is equivalent to  (\ref{2.6aa}).

Let now $\beta_0=0$. From (\ref{1.17}) and (\ref{2.9}) it follows that $\Delta_h\beta/h\to 0$, as $h\to 0$, i.e. $\dot\beta_0=0$. Hence, (\ref{2.6aa}) is also holds.

If $\beta_0=1$, then in an analogous way from (\ref{1.18}) we obtain that $\dot\beta_0=0$.
$\Box$
\begin{remark}\label{r1} \rm
In the case of i.i.d. observations which follow a distribution from a regular family, it is easy to see from the proof of
Theorem \ref{t1} that
\begin{equation}\label{2.9b}
  (\dot\beta_{\theta_0}(\psi,\phi))^2\leq \beta_{\theta_0}(\psi,\phi)(1-\beta_{\theta_0}(\psi,\phi))I(\theta_0)E_{\theta_0}\tau_\psi,
\end{equation}
where $I(\theta_0)$ is the Fisher information.
It is very likely that the same inequality  holds for a wide class of continuous-time stochastic processes
(as, for example, for  the class of processes with stationary and independent increments conside-red in \cite{roters} in relation with the locally most powerful tests). It is interesting to note that for the Wiener process with a linear drift it is shown in \cite{novikov06} that for the most powerful test $(\tau,\delta)$ with the type-I error probability equal to $\alpha$, it holds $\dot\beta_{\theta_0}(\tau,\delta)/\sqrt{E_{\theta_0}\tau}=\sqrt{\alpha(1-\alpha)}$, i.e. there is an equality in (\ref{2.9b}). It follows from (\ref{2.9b}) that if $\alpha\leq 0.5$, then for all $(\tau^\prime,\delta^\prime)$ such that $\beta_{\theta_0}(\tau^\prime,\delta^\prime)\leq\alpha$ and $E_{\theta_0}\tau^\prime\leq E_{\theta_0}\tau$ it holds $\dot \beta_{\theta_0}(\tau^\prime,\delta^\prime)\leq \dot \beta_{\theta_0}(\tau,\delta)$, i.e. the test $(\tau,\delta)$ is locally most powerful in a wider, than in \cite{roters}, class of sequential tests (in \cite{roters}, the class of tests $(\tau^\prime,\delta^\prime)$ such that $\beta_{\theta_0}(\tau^\prime,\delta^\prime)=\alpha$ and $E_{\theta_0}\tau^\prime\leq E_{\theta_0}\tau$ is considered). For the discrete-time processes of general form, the same extension of the class of tests is adopted in \cite{novikov09c}. We conjecture that,  under the conditions of \cite{roters}, this extension can be obtained in many cases, as easily as above, from  the corresponding generalization of (\ref{2.9b}) to the continuous-time case.
\end{remark}
\begin{theorem}\label{t2} Let Assumptions 1 to 3 be fulfilled. Then the power function $\beta(\psi,\phi)$ of every sequential test $(\psi,\phi)$ such that $E_{\theta_0}\tau_\psi<\infty$
is differentiable at $\theta=\theta_0$, and
\begin{equation}\label{2.3}
\dot\beta_{\theta_0}(\psi,\phi)=\sum_{n=1}^\infty E_{\theta_0}\left(s_n^\psi\phi_{n}\sum_{j=1}^n q_j\right),
\end{equation}
where
$$
q_n=q_n(x_n)=\frac{\dot f_{\theta_0,n}(x_n)}{f_{\theta_0,n}(x_n)}.
$$
\end{theorem}

{ P r o o f.}
Let $(\psi,\phi)$ be any sequential test such that $E_{\theta_0}\tau_\psi<\infty$. Let us prove that
\begin{equation}\label{2.10}
    (\beta_{\theta}(\psi,\phi)-\beta_{\theta_0}(\psi,\phi))/(\theta-\theta_0)-\sum_{n=1}^\infty E_{\theta_0}\left(s_n^\psi\phi_{n}\sum_{j=1}^n q_j\right)\to 0,\quad \theta\to\theta_0,
\end{equation}
that is,
\begin{equation}\label{2.11}
    \sum_{n=1}^\infty \int s_n^\psi \phi_n \left((f_{\theta}^n-f_{\theta_0}^n)/(\theta-\theta_0)-\dot f_{\theta_0}^n\right)d\mu^n\to 0,\quad \theta\to \theta_0,
\end{equation}
where $\dot f_{\theta_0}^n=(\sum_{j=1}^n q_j)f_{\theta_0}^n$ (it is not difficult to see that
$$E_{\theta_0}s_n^\psi\phi_{n}\sum_{j=1}^n q_j=\int  s_n^\psi \phi_n \dot f_{\theta_0}^nd\mu^n,
$$
because  from Assumption 2 it follows that   $\dot f_{\theta_0,j}=0$ $\mu$-almost everywhere on \\ $\{x:f_{\theta_0,j}(x)=0\}$).

From Assumption 2 it is not difficult to deduce that for any fixed $k\geq 1$
\begin{equation}\label{2.12}
    \sum_{n=1}^k \int s_n^\psi \phi_n \left((f_{\theta}^n-f_{\theta_0}^n)/(\theta-\theta_0)-\dot f_{\theta_0}^n\right)d\mu^n\to 0,\quad \theta\to \theta_0
\end{equation}
(practically it is differentiability of the product $f_\theta^n=\prod_{j=1}^nf_{\theta,j}$ in $L_1(\mu^n)$ under the condition of differentiability of  $f_{\theta,j}$ in $L_1(\mu)$). Because of that (\ref{2.11}) will follow if we prove that for every $\epsilon>0$ there exists $k>1$ such that
\begin{equation}\label{2.13}
    \limsup_{\theta\to\theta_0}|\sum_{n=k}^\infty \int s_n^\psi \phi_n \left((f_{\theta}^n-f_{\theta_0}^n)/(\theta-\theta_0)-\dot f_{\theta_0}^n\right)d\mu^n|<2\epsilon.
\end{equation}
Obviously, (\ref{2.13}) will follow if we show that such $k$ can be found that
\begin{equation}\label{2.14}
   \limsup_{\theta\to\theta_0} | \sum_{n=k}^\infty \int s_n^\psi \phi_n (f_{\theta}^n-f_{\theta_0}^n)/(\theta-\theta_0)d\mu^n|<\epsilon,
\end{equation}
 and
\begin{equation}\label{2.15}
    \sum_{n=k}^\infty \int s_n^\psi | \dot f_{\theta_0}^n|d\mu^n=\sum_{n=k}^\infty   E_{\theta_0}\left(s_n^\psi |\sum_{j=1}^n q_j|\right)<\epsilon.
\end{equation}

Let us turn first to the proof of (\ref{2.15}). To this end, let us note that by virtue of Lemma \ref{l4},
\begin{equation}\label{2.16}
    \sum_{n=1}^\infty   E_{\theta_0}\left(s_n^\psi \sum_{j=1}^n |q_j|\right)=\sum_{j=1}^\infty E_{\theta_0}|q_j|P_{\theta_0}(\tau_\psi\geq j),
\end{equation}
where the series on the right-hand side is finite, because it follows from Assumption 3 that
$ E_{\theta_0}|q_j|\leq{\gamma_2}<\infty. $

Hence, the series on the left-hand side of (\ref{2.16}) is converging, thus (\ref{2.15}) follows.

Let us prove now that there exists such $k$ that (\ref{2.14}) holds.
To this end, let us apply Lemma \ref{l1} with $G(x)=-\ln(x)$, $a_n=\phi_nI_{\{n\geq k\}}$,
$b_n=f_{\theta}^n/f_{\theta_0}^n$.

Let, for brevity,
$$\alpha_k= \sum_{n=k}^\infty E_{\theta_0}s_n^\psi\phi_n,\quad \alpha_k(\theta)= \sum_{n=k}^\infty E_{\theta}s_n^\psi\phi_n,$$
and let us suppose first that  $0<\alpha_k<1$. Then
\begin{eqnarray}\nonumber
 I(\theta_0,\theta;\psi)&=&\alpha_k\frac{\sum_{n=1}^\infty E_{\theta_0}s_n^\psi \phi_n I_{\{n\geq k\}} (-\ln(b_n))}{\alpha_k}\\\label{2.17}
 &+&(1-\alpha_k)\frac{\sum_{n=1}^\infty E_{\theta_0}s_n^\psi(1- \phi_nI_{\{n\geq k\}}) (-\ln(b_n))}{1-\alpha_k}.
\end{eqnarray}
Applying Lemma \ref{l1} to both fractions on the right-hand side of (\ref{2.17})  (as in the proof of (\ref{1.16})) we obtain
\begin{equation}\label{2.18}
 I(\theta_0,\theta;\psi)\geq -\alpha_k\ln\left(1+\frac{\alpha_k(\theta)-\alpha_k}{\alpha_k}\right)-(1-\alpha_k)\ln\left(1-\frac{\alpha_k(\theta)-\alpha_k}{1-\alpha_k}\right).
\end{equation}
Because, according to (\ref{2.9}), the left-hand side of (\ref{2.18}) tends to zero as $\theta\to\theta_0$,  in complete analogy with the proof of Theorem \ref{t1}, we first get that $\alpha_k(\theta)\to\alpha_k$, as $\theta\to\theta_0$, and then, applying the Taylor formula for $\ln(1+x)$ at $x=0$ up to the second-order terms:
$$
\frac{(\alpha_k(\theta)-\alpha_k)^2}{2\alpha_k(1-\alpha_k)}+o((\alpha_k(\theta)-\alpha_k)^2)\leq \gamma_1 (\theta-\theta_0)^2.
$$
Therefore,
$$
\limsup_{\theta\to\theta_0}\left|\frac{\alpha_k(\theta)-\alpha_k}{\theta-\theta_0}\right|\leq \sqrt{2\gamma_1\alpha_k}\leq \sqrt{2\gamma_1P_{\theta_0}(\tau_\psi\geq k)}.
$$
Because of that, (\ref{2.14}) follows if $\sqrt{2\gamma_1P_{\theta_0}(\tau_\psi\geq k)}\leq \epsilon$, which can be done, since, by condition,  $E_{\theta_0}\tau_\psi<\infty$.

Let us consider now the case $\alpha_k=\sum_{n\geq k}E_{\theta_0}s_n^\psi\phi_n=0$. By Lemma \ref{l1}
$$
I(\theta_0,\theta;\psi)=\sum_{n=1}^\infty E_{\theta_0}s_n^\psi (-\ln\frac{f_\theta^n}{f_{\theta_0}^n})(1-\phi_nI_{\{n\geq k\}})
$$
$$
\geq -\ln ( \sum_{n=1}^\infty E_{\theta_0}s_n^\psi
\frac{f_\theta^n}{f_{\theta_0}^n}(1-\phi_nI_{\{n\geq k\}}))\geq -\ln(1-\sum_{n=k}^\infty E_\theta \phi_nI_{\{n\geq k\}})
$$
$$=-\ln(1-\alpha_k(\theta))\geq \alpha_k(\theta)=\alpha_k(\theta)-\alpha_k.
$$
By virtue of (\ref{2.9}) it follows from this that $$\lim_{\theta\to\theta_0}\frac{\alpha_k(\theta)-\alpha_k}{|\theta-\theta_0|}=0,$$
that is (\ref{2.14}) holds also in this case.

 Analogously it can be proved that  if $\alpha_k=1$, then
$$\lim_{\theta\to\theta_0}\frac{1-\alpha_k(\theta)}{|\theta-\theta_0|}=0,$$
that is  (\ref{2.14}) holds as well.
$\Box$
\begin{remark}\label{r3}\rm Theorem \ref{t2} is a generalization, to the case of non-identically distributed
observations and of randomized stopping- and decision rules, of Lemma 4.1.4 \cite{irle}. For i.i.d. observations
this result was announced in \cite{berk} and ascends to the unpublished work \cite{abraham}. The proof of this result in  \cite{irle} follows \cite{mueller86}.
Similar questions about the existence of the second derivatives of the power function of sequential tests apparently remain not answered until now (see \cite{mueller-funk}).
\end{remark}

{\bf  4. The structure of optimal sequential tests. Truncated stopping rules.  }
In this section we characterize the optimal  sequential tests that take, at most,  some fixed number $N$  observations.

  For any natural $N$ let us denote by $\mathscr F^N$ the class of {\em truncated} (at $N$) stopping rules, i.e. such $\psi$ that $\psi_N\equiv 1$.

Let us start the construction with defining the following functions.

Let $g(z)=\min\{0,z\}$, $z\in\mathbb R$. Let us define for all $N\geq 1$
 and $n=1,\dots,N$ the functions $v_n^N(z)=v_n^N(z;c)$, $z\in\mathbb R$, starting from
 \begin{equation}\label{3.1a}
    v_N^N(z)\equiv g(z),\; z\in\mathbb R,
 \end{equation}
by means of the following recurrent relations
\begin{equation}\label{3.5}
v_{n-1}^N(z;c)=\min
\left\{g(z), c+E_{\theta_0}v_n^N\left(
z-q_n;c\right)\right\},
\end{equation}
$n=N,N-1,\dots,1$, where, by definition, $q_n=q_n(x_n)=\dot f_{\theta_0,n}(x_n)/ f_{\theta_0,n}(x_n)$. Let
\begin{equation}\label{3.2a}
r_{n-1}^N(z)=r_{n-1}^N(z;c)=E_{\theta_0}v_n^N\left(
z-q_n;c\right),
\end{equation}
$n=1,2,\dots,N$.

For any  $b\in\mathbb R$ and $c>0$ define, following \cite{novikov09c}, the ``Lagrange-multiplier function''
\begin{equation}\label{4.2}
L_N(\psi;b,c)=\sum_{n=1}^NE_{\theta_0}s_n^\psi\left(nc+\min\left\{0,b-\sum_{i=1}^nq_i\right\}\right)
\end{equation}
for all $\psi\in \mathscr F^N$ (see (4.2) in \cite{novikov09c}).

Let also
$$z_n=z_n(x_1,\dots,x_n)=\sum_{i=1}^n q_i(x_i)$$
(if $\prod_{i=1}^nf_{\theta_0,i}(x_i)=0$, let us suppose that  $z_n=0$).
\begin{theorem} Suppose that Assumption 2 is fulfilled.

Then for all $\psi\in\mathscr F^N$
\begin{equation}\label{4.2a}
    L_N(\psi;b,c)\geq c+r_0^N(b;c).
\end{equation}
The equality in (\ref{4.2a}) is attained if and only if
\begin{equation}\label{4.5}
I_{\{g(b-z_n)< c+r_n^N(b-z_n;c)\}}\leq\psi_{n}\leq I_{\{g(b-z_n)\leq c+r_n^N(b-z_n;c)
\}}
\end{equation}
$\mu^n$-almost everywhere on $T_n^\psi\cap\{f_{\theta_0}^n>0\}$ for all $n=1,2,\dots, N-1$.
\end{theorem}

{ P r o o f.} It is sufficient to express the elements of the optimal stopping rule from Corollary 4.1 \cite{novikov09c}
  ($V_n^N$ and $R_n^N$) through the corresponding functions $v_n^N$ and $r_n^N$.
Let us show that for all  $N=1,2,\dots$ and $n\leq N$
\begin{equation}\label{4.5a}
    V_n^N = v_n^N(b-z_n) f_{\theta_0}^n
\end{equation}
$\mu^n$-almost everywhere.

Let us conduct the proof by induction over $n=N,N-1,\dots,1$.
All equalities between functions of observations $(x_1,\dots,x_n)$ will be understood $\mu^n$-almost everywhere.

For $n=N$, obviously,
$$V_N^N=l_N=\min\{0, b-z_N\}f_{\theta_0}^N=v_N^N(b-z_N)f_{\theta_0}^N.$$
Let us suppose that  (\ref{4.5a}) is fulfilled for some $n\leq N$. Then
$$
V_{n-1}^N=\min\{l_{n-1},cf_{\theta_0}^{n-1}+\int V_n^Nd\mu(x_n)\}$$
$$=\min\left\{\min\{0,bf_{\theta_0}^{n-1}-\dot f_{\theta_0}^{n-1}\},cf_{\theta_0}^{n-1}+\int v_n^N(b-z_n) f_{\theta_0}^nd\mu(x_n) \right\}
$$
$$
=\min\left\{g(b-z_{n-1}),c+\int v_n^N\left(b-z_{n-1}-q_n\right) f_{\theta_0,n}(x_n)d\mu(x_n)\right\}f_{\theta_0}^{n-1}
$$
$$
=v_{n-1}^N(b-z_{n-1})f_{\theta_0}^{n-1}.
$$
Thus, (\ref{4.5a}) is proved.

 We have now $$R_{n-1}^N=\int V_{n}d\mu(x_{n})=\int v_{n}^N \left(b-z_{n-1}-q_n\right)f_{\theta_0,n}(x_n)d\mu(x_n)f_{\theta_0}^{n-1}$$
$$
=r_{n-1}^N(b-z_{n-1})f_{\theta_0}^{n-1}
$$
for all $n=1,2,\dots,N$.

It is obvious now that  (\ref{4.5}) is equivalent to  (4.5) in \cite{novikov09c}, if $f_{\theta_0}^n>0$.
$\Box$

\begin{corollary}\label{c1} Let us suppose that Assumption 2 is fulfilled, and let $b>0$ is any real number.

 Let $\psi\in \mathscr F^N$ be any stopping rule satisfying (\ref{4.5}) $\mu^n$-almost everywhere on $T_n^\psi$ for all $n=1,2,\dots, N-1$, and let the decision rule $\phi$ be such that
\begin{equation}\label{1.7}
I_{\{z_n>b\}}\leq\phi_n\leq I_{\{z_n\geq b\}}
\end{equation}
$\mu^n$-almost everywhere on $S_n^\psi$ for all $n=1,2,\dots,N$.

 Then the test $(\psi,\phi)$ is locally most powerful in the class of all (truncated) tests $(\psi^\prime,\phi^\prime)$ с $\psi^\prime\in\mathscr F^N$, in the sense that
\begin{equation}\label{1.8}
    \dot \beta_{\theta_0}(\psi,\phi)\geq \dot \beta_{\theta_0}(\psi^\prime,\phi^\prime)
\end{equation}
whenever
\begin{equation}\label{1.9}
    \alpha(\psi^\prime,\phi^\prime)\leq \alpha(\psi,\phi)\quad\mbox{and}\quad {\mathscr N}_{\theta_0}(\psi^\prime)\leq {\mathscr N}_{\theta_0}(\psi).
\end{equation}
The inequality in  (\ref{1.8}) is strict, if at least one of the inequalities in (\ref{1.9}) is strict.
If in all inequalities in (\ref{1.8}) and (\ref{1.9}) the equalities are attained, then $\psi^\prime$ also satisfies (\ref{4.5}) $\mu^n$-almost everywhere on $T_n^{\psi^\prime}$ for all $n=1,2,\dots, N-1$ (with $\psi_n^\prime$ instead of $\psi_n$), and $\phi^\prime$ satisfies (\ref{1.7}) (with $\phi_n^\prime$ instead of $\phi_n$) $\mu^n$-almost everywhere on $S_n^{\psi^\prime}$ for all $n=1,2,\dots,N$.
\end{corollary}

A more detailed description of optimal stopping rules can be obtained from the investigation of properties of all functions involved in (\ref{4.5}). Let us formulate the corresponding properties in the following lemmas.

\begin{lemma}\label{l5}  The functions $v_n^N(z)$, $n=0,\dots,N$, $N=1,2,\dots$ defined by (\ref{3.5})
possess the following properties:\\
1) $v_n^N(z)\leq g(z),\;z\in\mathbb R$,\\
2) $v_n^N(z)$  is a concave and continuous function on $\mathbb R$,\\
3) $v_n^N(z)$ is a non-decreasing function on $\mathbb R$,\\
4) $z-v_n^N(z)$ is a non-decreasing function on $\mathbb R$,\\
5) $g(z)-v_n^N(z)\to 0$ as $z\to\pm\infty $.
\end{lemma}

{ P r o o f.} We will need the following simple lemma in the proof of this, and some subsequent, lemmas.
\begin{lemma}\label{l6}
Let $F$ be a concave function on $\mathbb R$. Then for all $n\geq 1$
$$G_n(z)=E_{\theta_0}F\left(z-q_n\right)$$
is a concave function of $z$. In addition, $G_n(z)\leq F(z)$, $z\in\mathbb R$.
\end{lemma}
Property 1) is a direct consequence of definitions (\ref{3.1a}) and (\ref{3.5}).

We prove properties  2) to 5) simultaneously, using induction over $n=N,N-1,\dots,1$.

For $v_N^N(z)\equiv g(z)$ all the properties mentioned in 2) -- 5) are obvious.

Let us suppose that properties  2) -- 5) hold for some $n\leq N$. Let us prove that they also hold for $v_{n-1}^N$.

 By virtue of (\ref{3.5}), $v_{n-1}^N$ is a minimum of two concave functions (the second one is concave by Lemma \ref{l6}).
Thus, $v_{n-1}^N$ is also concave.

Now it follows from Theorem 10.1 \cite{rockafellar} that  $v_{n-1}^N$ is continuous.

 If $v_n^N(z)$ is non-decreasing, then by (\ref{3.5})
$v_{n-1}^N(z)$ is also non-decreasing.
Because $z-v_n^N(z)$ is non-decreasing, we have
$$ z-v_{n-1}^N(z)
= \max\left\{\max\{0,z\}, -c+E_{\theta_0}\left((z-q_n)-v_n^N\left(z-q_n\right)\right)\right\} $$
is non-decreasing as well, since the mathematical expectation on the right-hand side is a non-decreasing function of $z$.

Let us finally show that $g(z)-v_{n-1}^N(z)\to 0$, as $z\to\pm\infty$ (property 5) of the lemma).

Let first $z_k$, $k=1,2,\dots$, be a monotone increasing sequence, $z_k\to\infty$, $k\to\infty$.

For  $k$ large enough, $z_k>0$, thus, for such $k$, $g(z_k)=0$, so that
$$ g(z_k)-v_{n-1}^N(z_k)
= -\min\left\{0, c+E_{\theta_0}v_n^N\left(z_k-q_n\right)\right\}\to 0, $$
 as $k\to\infty$, because the mathematical expectation converges to zero by the Lebesgue's dominated convergence theorem. Indeed, by the supposition of the induction, $v_n^N(z_k-q_n)\to 0$, as $k\to\infty$, and $$v_n^N(z_1-q_n)\leq v_n^N(z_k-q_n)\leq 0.$$ Here the function $v_n^N(z_1-q_n)$ is integrable, because by virtue of properties 3) and 4) we have: $$0\leq g(z)-v_n^N(z)\leq -v_n^N(0)<\infty,$$
so $$v_n^N(z_1-q_n)\geq g(z_1-q_n)+v_n^N(0),$$
 and, in addition, $E_{\theta_0}|g(z_1-q_n)|\leq E_{\theta_0}|z_1-q_n|<\infty$.

Let now $z_k$, $k=1,2,\dots$, be a monotone decreasing sequence, $z_k\to-\infty$, $k\to\infty$.
For $k$ sufficiently large $z_k<0$,  so $g(z_k)=z_k$, and
$$ g(z_k)-v_{n-1}^N(z_k) = -\min\left\{0,
c-E_{\theta_0}\left(
\left(z_k-q_n\right)
-v_n^N\left(z_k-q_n\right)
\right) \right\}\to 0 $$
 as $z\to\infty$, because the mathematical expectation converges to zero by the Lebesgue dominated convergence theorem. Indeed, $\left(z_k-q_n\right)
-v_n^N\left(z_k-q_n\right)\to 0$, as $k\to\infty$, by virtue of property 5), and in addition
$$
\left(z_k-q_n\right)
-v_n^N\left(z_k-q_n\right)\leq \left(z_1-q_n\right)
-v_n^N\left(z_1-q_n\right)
$$
by virtue of property  4), where the function on the right-hand side of the inequality is integrable, for the same reasons as above.
$\Box$

\begin{lemma}\label{l7}  The functions $r_n^N(z)$, $n=0,\dots,N$, $N=1,2,\dots$, defined by (\ref{3.2a}),
possess the following properties:\\
1) $r_{n}^N(z)\leq v_n^N(z),\;z\in\mathbb R$, \\
2) $r_n^N(z)$ as a function of $z\in\mathbb R$ is concave and continuous,\\
3) $r_n^N(z)$ as a function of  $z\in\mathbb R$ is non-decreasing,\\
4) $z-r_n^N(z)$ as a function of  $z\in\mathbb R$ is non-decreasing,\\
5) $g(z)-r_n^N(z)\to 0$,  as $z\to\pm\infty $.
\end{lemma}

{ P r o o f.} 1)
We have by definition:
$$ r_n^N(z)-v_n^N(z)=-\min\{g(z)-r_n^N(z),c\}$$
$$ \leq-\min \left\{E_{\theta_0}\left(
g\left(z-q_n\right)
-v_n^N\left(z-q_n\right)
\right),\, c\right\}\leq 0$$
where the first inequality follows from the Jensen inequality, and the second from property 1) of Lemma \ref{l5}.

2) By virtue of property  2) of Lemma \ref{l5}, $v_{n+1}^N(z-q_{n+1}) $ is a concave function of $z$. By Lemma \ref{l6}, the concavity of $r_n^N$ follows from this. The continuity of $r_n^N$ follows now from Theorem 10.1 \cite{rockafellar}.

3) By virtue of property 3) of Lemma \ref{l5},
$v_{n+1}^N(z-q_{n+1}) $ is a non-decreasing function of $z$, it follows from this that $r_n^N(z)=E_{\theta_0}v_{n+1}^N(z-q_{n+1})$ is a non-decreasing function of $z$.

4) In the same way $z-r_n^N(z)=E_{\theta_0}((z-q_{n+1})-v_{n+1}^N(z-q_{n+1}))$ is a non-decreasing function of $z$.

5) See the proof of property 5) of Lemma \ref{l5}.
$\Box$

\begin{lemma}\label{l8} If  $c+r_n^N(0)\leq 0$, then in each region $\{z\leq 0\}$ and $\{z\geq 0\}$ there exists a unique solution to the equation
\begin{equation}\label{4.6} c+r_n^N(z)=g(z),\end{equation} that will be denoted  $A_n^N=A_n^N(c)\leq 0$ and $B_n^N=B_n^N(c)\geq 0$. In addition, $g(z)>c+r_{n}^N(z)$
if and only if  $A_n^N<z<B_n^N$.

If $c+r_n^N(0)>0$, then the equation (\ref{4.6}) does not have a solution.
\end{lemma}

{ P r o o f.}
The function $g(z)-r_n^N(z)$ is continuous by property 2) of Lemma \ref{l7}, and non-negative by property 1) of Lemma \ref{l7} and property 1) of Lemma \ref{l5}.

 By virtue of properties 3) and 4) of Lemma \ref{l7}, $g(z)-r_n^N(z)$ is non-decreasing for $z\leq 0$  and non-increasing for $z\geq 0$. Hence, its maximum value is attained at $z=0$ and is equal to $-r_n^N(0)$, so that for $c+r_n^N(0)>0$ the equation (\ref{4.6}) can not have a solution.

Let us prove that otherwise there is a unique solution to the equality  (\ref{4.6}) for $z\leq 0$ and for $z\geq 0$. For example, let us prove this for $z\leq 0$ -- the other case is completely analogous.

 For $z\leq 0$ the function $g(z)-r_n^N(z)=z-r_n^N(z)$ is convex, continuous, non-decreasing, and such that $g(z)-r_n^N(z)\to 0$, as $z\to-\infty$ (Lemma \ref{l7}). It is easy to see that any function on $(-\infty,0]$ with this properties  takes any positive value not exceeding its maximum value, and does so only once.  Because, by supposition, $0<c\leq -r_n^N(0)=\max_{z\leq 0}\{g(z)-r_n^N(z)\}$, it follows from this that for $z\leq 0$ there is a unique solution to  $g(z)-r_n^N(z)=c$, $A_n^N$. It addition, it is obvious that for  $z>A_n^N$ it holds $g(z)-r_n^N(z)>c$, that is, $g(z)>c+r_n^N(z)$. The latter inequality is satisfied only if $z>A_n^N$, because, by the monotonicity, $g(z)-r_n^N(z)\leq c$ for all $z\leq A_n^N$.
$\Box$

If  $c+r_n^N(0)\leq 0$, let us denote by $\Delta_n^N$ the interval $(A_n^N,\,B_n^n)$ and by  $\bar \Delta_n^N$ the closed interval $[A_n^N,\,B_n^n]$. If $c+r_n^N(0)> 0$, then let, by definition,  $\bar \Delta_n^N=\Delta_n^N=\emptyset$. Note that $\Delta_n^N=\Delta_n^N(c)$ and $\bar\Delta_n^N=\bar\Delta_n^N(c)$.

\begin{corollary}\label{c2}
Under the conditions of Corollary \ref{c1}
its assertion remains true after substituting all the references to (\ref{4.5}) for
the references to
\begin{equation}\label{4.7}
\begin{array}{c}
  I_{\{b-z_n\in\Delta_n^N(c)\}}\leq 1-\psi_{n}\leq I_{\{b-z_n\in\bar\Delta_n^N(c)
\}}.
\end{array}
\end{equation}
\end{corollary}

{ P r o o f.}
From Lemma  \ref{l8} it follows that $g(b-z_n)> c+r_n^N(b-z_n;c)$ if and only if
$b-z_n\in\Delta_n^N(c)$, and
$g(b-z_n)\geq c+r_n^N(b-z_n;c)$ if and only if
$b-z_n\in\bar\Delta_n^N(c)$. Therefore, (\ref{4.7}) is equivalent to (\ref{4.5}).
$\Box$\vspace{5mm}

{\bf  5. The structure of optimal sequential tests. The general case.}
In this section we characterize the structure of optimal sequential tests when there is no restriction
on the maximum number of observations.

The idea of what follows is to let the maximum number of observations $N$ we supposed fixed in the previous section, tend to infinity. Doing this, we prove the convergence of all elements defining the structure of optimal rules in the truncated problem to the corresponding elements in the non-truncated problem (see \cite{novikov09c}).

Let us start with the following lemma.
\begin{lemma}\label{l9}
For all $N\geq 1$ and $n\leq N$ \\
1) $ v_n^N(z)\geq v_n^{N+1}(z)$,\\
2) $ r_n^N(z)\geq r_n^{N+1}(z)$\\
for all $z\in\mathbb R$.
\end{lemma}

{ P r o o f.}
Let us prove inequality 1) by induction over $n=N,N-1,\dots, 1$.
Let $n=N$. Then
$$ v_N^{N+1}(z) = \min\{g(z), c+E_{\theta_0} v_{N+1}^{N+1}(z-q_n)\}
\leq g(z) = v_{N+1}^{N+1}(z). $$
Let us suppose that the inequality $v_n^N\geq v_n^{N+1}$ is fulfilled for some $n$, $N\geq n>1$.
Then
$$ v_{n-1}^N(z) = \min\{g(z),c+E_{\theta_0}v_n^N(z-q_n)\}
\geq \min\{g(z),c+E_{\theta_0}v_n^{N+1}(z-q_n)\}=v_{n-1}^{N+1}. $$
Thus, the inequality is also fulfilled for $n-1$ which completes the induction.

Assertion 2) is a direct consequence of assertion 1)
by virtue of (\ref{3.2a}).
$\Box$

Because, by Lemma \ref{l9}, $v_n^N(z)$ and $r_n^N(z)$ are non-increasing  with respect to $N$ for each $z\in\mathbb R$,
there exist the limits  (finite or not)
\begin{equation}\label{4.13}
v_n(z)=v_n(z;c)=\lim_{N\to\infty}v_n^N(z;c),
\end{equation}
\begin{equation}\label{4.14}
r_n(z)=r_n(z;c)=\lim_{N\to\infty}r_n^N(z;c).
\end{equation}
In addition, passing to the limit as $N\to\infty$ in (\ref{3.5}) and (\ref{3.2a}), for $n=1,2,\dots$, we get:
\begin{equation}\label{4.15}
v_{n-1}(z;c)=\min
\left\{g(z), c+E_{\theta_0}v_n\left(
z-q_n;c\right)\right\},
\end{equation}
\begin{equation}\label{4.16}
r_{n-1}(z;c)=E_{\theta_0}v_n\left(z-q_n;c\right).
\end{equation}

Let us define $\mathscr F$ as the class of stopping rules with finite average sample number under the null-hypothesis:
$$ \mathscr F=\{\psi:\, E_{\theta_0}\tau_\psi<\infty\}. $$

Let us show that, under Assumptions 1 -- 3, for each $\psi\in\mathscr F$ it holds $L_N(\psi;b,c)\to L(\psi;b,c)$, $N\to\infty$.

\begin{lemma}\label{l2} Let Assumptions 1 -- 3 are fulfilled and let $\psi\in\mathscr F$. Then
$$
L_N(\psi;b,c)\to L(\psi;b,c),
$$
as $N\to\infty$ for all $c>0$ and $b\in\mathbb R$.
\end{lemma}

{P r o o f.} Completely analagous to the proof of Lemma 4.4 in \cite{novikov09c}, with the only difference that in order to prove
\begin{equation}\label{4.16a}
    \int t_N^\psi l_N d\mu^N\to 0, \quad N\to \infty,
\end{equation}
we can use   in the case of independent observations, instead of Assumption  3 \cite{novikov09c},
a weaker Assumption 3.
Indeed, in terms of this article
\begin{eqnarray}
 \nonumber \int t_N^\psi |l_N| d\mu^N&=&E_{\theta_0}t_N^\psi|\min\{0,b-\sum_{j=1}^Nq_j\}|\leq E_{\theta_0}t_N^\psi|b-\sum_{j=1}^Nq_j|\\
   &\leq& |b|P_{\theta_0}(\tau_\psi\geq N)+E_{\theta_0}t_N^\psi\sum_{j=1}^N|q_j|\label{4.16b}
\end{eqnarray}
The first summand on the right-hand side of (\ref{4.16b}) tends to zero as $N\to\infty$ by the condition $E_{\theta_0}\tau_\psi<\infty$. To prove the fact that the second summand on the right-hand side of (\ref{4.16b}) also tends to zero,let us note that it follows from Assumption 3 that  the series on the right-hand side of  (\ref{2.16}) is finite, and hence so is the left-hand side,  thus
\begin{equation}\label{4.16c}
    \sum_{n=N}^\infty E_{\theta_0}s_n^\psi \sum_{j=1}^N |q_j|\leq \sum_{n=N}^\infty E_{\theta_0}s_n^\psi \sum_{j=1}^n |q_j|\to 0
\end{equation}
as $N\to\infty$. Since $E_{\theta_0}\sum_{j=1}^N |q_j|<\infty$, we easily get from this that
$$
\sum_{n=N}^\infty E_{\theta_0}s_n^\psi \sum_{j=1}^N |q_j|=E_{\theta_0}t_N^\psi \sum_{j=1}^N |q_j|\to 0
$$
as $N\to \infty$. $\Box$

By virtue of Lemma \ref{l2} we can pass to the limit on both sides of the inequality in (\ref{4.2a}), so
$$
L(\psi;b,c)\geq c+r_0(b;c)
$$
for all $\psi\in \mathscr F$, if Assumptions 1 to 3 are fulfilled.
In addition, by Lemma 4.3 in \cite{novikov09c}, $\inf_{\psi\in\mathscr F}L(\psi;b,c)=c+r_0(b;c)$.

Let us show that under Assumptions 1 -- 3 the problem of minimization of $L(\psi;b,c)$ is finite (in terms of \cite{novikov09c}), more precisely, that the following lemma holds.

\begin{lemma}\label{l10}
If Assumptions 1 to 3 are fulfilled, and let $b>0$, $c>0$ be any real numbers. Then for all $\psi\in \mathscr F $
\begin{equation}\label{4.4}
L(\psi;b,c)\geq -\frac{\gamma_1}{8c}
\end{equation}
\end{lemma}

{ P r o o f}. It follows from Theorem \ref{t1} that
$$
\dot\beta_{\theta_0}(\psi,\phi)\leq \sqrt{\frac{\gamma_1}{2} E_{\theta_0}\tau_\psi},
$$
so
$$
L(\psi,\phi;b,c)\geq cE_{\theta_0}\tau_\psi-\sqrt{\frac{\gamma_1}{2} E_{\theta_0}\tau_\psi}\geq -\frac{\gamma_1}{8c},
$$
from which (\ref{4.4}) follows, because, by virtue of  Corollary 3.1 in \cite{novikov09c},
$$L(\psi;b,c)=\inf_{\phi}L(\psi,\phi;b,c).$$$\Box$

\begin{remark}\label{r2}
It follows from Lemma  \ref{l10} that $$\inf_{\psi\in\mathscr F}L(\psi;b,c)=c+r_0(b;c)\geq  -\frac{\gamma_1}{8c}>-\infty$$ for all $b>0$ and $c>0$.

This also implies that  $c+r_n(b;c)>-\frac{\gamma_1}{8c}$ for all $b>0$, $c>0$ and all $n\geq 0$. Indeed, by construction, $r_n$ is ``the $r_0$ function'' for the problem of testing $H_0:\theta=\theta_0$ vs. $H_1:\, \theta>\theta_0$ about the parameter of distribution of the process $X_1,X_2,\dots$ for which $X_1\sim f_{\theta,n+1},\, X_2\sim f_{\theta,n+2}, \dots$.
\end{remark}
Now Theorem 4.2 \cite{novikov09c} takes the following form.

\begin{theorem}\label{t3}
Suppose that Assumption 1 to 3 are fulfilled.

If there is a $\psi\in\mathscr F$ such that
\begin{equation}\label{4.4b}
L(\psi;b,c)=\inf_{\psi^\prime\in\mathscr F}L(\psi^\prime;b,c),
\end{equation}
then
\begin{equation}\label{4.5b}
I_{\{g(b-z_n)< c+r_n(b-z_n;c)\}}\leq\psi_{n}\leq I_{\{g(b-z_n)\leq c+r_n(b-z_n;c)
\}}
\end{equation}
$\mu^n$-almost everywhere on $T_n^\psi\cap\{f_{\theta_0}^n>0\}$ for all $n=1,2,\dots$.

Reversely, if a stopping rule $\psi$ satisfies (\ref{4.5b}) $\mu^n$-almost everywhere on $T_n^\psi\cap\{f_{\theta_0}^n>0\}$ for all $n=1,2,\dots$, and $\psi\in\mathscr F$, then it satisfies (\ref{4.4b}).
\end{theorem}
For the proof of Theorem \ref{t3} we need the following lemma.
\begin{lemma}\label{l11}  The functions $r_n(z)$, $n=0,\dots$ defined by (\ref{4.14}),
have the following properties:\\
1) $r_n(z)\leq v_n(z)\leq g(z),\;z\in\mathbb R$, \\
2) $r_n(z)$ as a function of $z\in\mathbb R$ is concave and continuous,\\
3) $r_n(z)$ as a function of $z\in\mathbb R$ is non-decreasing,\\
4) $z-r_n(z)$ as a function of  $z\in\mathbb R$ is non-decreasing,\\
5) $g(z)-r_n(z)\to 0$,  as $z\to\pm\infty $.
\end{lemma}

{ P r o o f.} Properties 1) -- 4) follow from the corresponding properties of Lemma \ref{l7} by passing to the limit as $N\to\infty$ (the continuity in property 2) follows from the concavity).

To prove property 5) it is sufficient to show that
$z-r_n(z)\to 0$ as $z\to-\infty$ and $r_n(z)\to 0$ as $z\to+\infty$.

To prove that $r_n(z)\to 0$, $z\to+\infty$,
it suffices to show, by virtue of (\ref{4.16}) and the monotone convergence theorem, that $v_n(z)\to 0$, $z\to +\infty$.

By property 3)  the limit $\lim_{z\to+\infty} v_n(z;c)=\lambda_n(c)$ (in what follows, briefly, $\lambda_n$)
exists for all $n=1,2,\dots$.
From (\ref{4.16}) it follows that $\lim_{z\to\infty}r_{n-1}(z,c)=\lambda_n(c)$, $n=1,2,\dots$.
Passing to the limit, as $z\to\infty$,
in (\ref{4.15}) we get that
\begin{equation}\label{4.17}
    \lambda_{n}=\min\{0,c+\lambda_{n+1}\}
\end{equation}
 for all$n=1,2,\dots$. From (\ref{4.17}) it is obvious that if for some $n\geq 1$ $\lambda_n<0$, then $\lambda_n=c+\lambda_{n+1}<0$, therefore, $\lambda_{n+1}=c+\lambda_{n+2}<0$, and so on for all other $n$. This immediately leads to a contradiction because then $\lambda_{n+1}=\lambda_n-c$, $\lambda_{n+2}=\lambda_{n+1}-c=\lambda_n-2c$, \dots $\lambda_{n+k}=\lambda_n-kc$, $\dots$, and consequently
$r_{n+k-1}(0;c)\leq \lambda_n-kc$ for all $k\geq 1$, which contradicts the fact that $r_{n+k-1}(0;c)\geq -\frac{\gamma_1}{8c}-c$ for all $k\geq 1$ (see Remark \ref{r2}).

Hence, $\lambda_n(c)=\lim_{z\to\infty}r_{n-1}(z;c)=0$ for all $n\geq 1$.

Let us consider now the case $z\to-\infty$. It is easy to see that
$$v_{n-1}^N(z;c)-z=\min\{\min\{0,-z\},c+E_{\theta_0}(v_n^N(z-q_n;c)-(z-q_n))\}$$
which entails, by passing to the limit as $N\to\infty$, that
\begin{equation}\label{4.18}
v_{n-1}(z;c)-z=\min\{\min\{0,-z\},c+E_{\theta_0}(v_n(z-q_n;c)-(z-q_n))\}
\end{equation}
where, by virtue of property 4) of Lemma \ref{l5},
the functions $v_{n}(z;c)-z$ are non-increasing for all $n=1,2,\dots$.
Being so, there exist limits $\lim_{z\to-\infty}v_{n}(z;c)-z=\lambda_n(c)\leq 0$ (let, for brevity, $\lambda_n=\lambda_n(c)$). In the same way as above, passing to the limit as $z\to -\infty$ in (\ref{4.18}), we get
$$
\lambda_{n}=\min\{0,c+\lambda_{n+1}\},
$$
$n=1,2,\dots$. Supposing again that $\lambda_n<0$, we obtain that
$\lambda_{n+k}=\lambda_n-kc\to -\infty$, as $k\to \infty$. Therefore, for all $z\leq 0$, $r_{n+k-1}(z;c)-z\leq \lambda_n-kc$ (by property 4) of Lemma \ref{l11}). In particular, putting $z=0$, we get that $r_{n+k-1}(0;c)\leq \lambda_n-kc$ for all $k=1,2,\dots$, which is   a contradiction, again,  with the fact that all $r_n(0;c)$ are bounded from below by the same constant, for all $n=0,1,2,\dots$.

Consequently, $\lambda_n=\lim_{z\to-\infty} (r_{n-1}(z;c)-z)=0$ for all $n=1,2,\dots$.
$\Box$

{ P r o o f}\,\, of Theorem \ref{t3}. The necessity immediately follows fromTheorem 4.2 \cite{novikov09c}.
To prove the sufficiency it is sufficient to show that
\begin{equation}\label{4.8}
   \int t_n^\psi (l_n-V_n)d\mu^n\to 0
\end{equation}
as $n\to\infty$ (see (4.16) в \cite{novikov09c}).

It follows from (\ref{4.5a}) that $V_n=v_n(b-z_n)f_{\theta_0}^n$. In addition, we know that $l_n=g(b-z_n)f_{\theta_0}^n$. Therefore, the integral in (\ref{4.8}) coincides with $$ \int t_n^\psi (l_n-V_n)d\mu^n=E_{\theta_0}t_n^\psi(g(b-z_n)-v_n(b-z_n))$$
\begin{equation}\label{4.9}
    \leq E_{\theta_0}t_n^\psi(g(b-z_n)-r_n(b-z_n))
\end{equation}
(the latter inequality is valid by  property 1) of Lemma \ref{l11}). By virtue of properties 3) and 4) of the same Lemma we have for all $z$
$$
0\leq g(z)-r_n(z)\leq -r_n(0)\leq \frac{\gamma_1}{8c}+c
$$
(we used Lemma \ref{l10} for the last estimation  (see Remark \ref{r2})).
Thus, from  (\ref{4.9}) it follows that
$$
0\leq\int t_n^\psi (l_n-V_n)d\mu^n\leq (\frac{\gamma_1}{8c}+c)P_{\theta_0}(\tau_\psi\geq n)\to 0
$$
as $n\to\infty$, because, by the condition of the theorem, $\psi\in\mathscr F$, and so $E_{\theta_0}\tau_\psi<\infty$.
$\Box$

The next theorem follows from Theorem  \ref{t3} with the help of Theorems 3.1 and 3.2 from \cite{novikov09c}, and gives a solution of the source conditional problem (see the Introduction) in the class of all sequential tests with stopping rules from $\mathscr F$.

\begin{theorem}\label{t4} Suppose that Assumptions 1--3 are fulfilled, and let $b>0$, $c>0$ be any real numbers.

Let $\psi$ be any stopping rule satisfying
 \begin{equation}\label{4.15a}
    I_{\{g(b-z_n)< c+r_n(b-z_n;c)\}}\leq\psi_{n}\leq I_{\{g(b-z_n)\leq c+r_n(b-z_n;c)
\}}
 \end{equation}
 $\mu^n$-almost everywhere on $T_n^\psi\cap\{f_{\theta_0}^n>0\}$ for all $n=1,2,\dots$, and let the decision rule $\phi$ be such that
\begin{equation}\label{1.7a}
I_{\{z_n>b\}}\leq\phi_n\leq I_{\{z_n\geq b\}}
\end{equation}
$\mu^n$-almost everywhere on $S_n^\psi\cap\{f_{\theta_0}^n>0\}$ for all $n=1,2,\dots,N$.

 Suppose that $\psi\in\mathscr F$ (i.e. $E_{\theta_0}\tau_\psi<\infty$).

Then the test $(\psi,\phi)$ is locally most powerful in the class of all tests $(\psi^\prime,\phi^\prime)$ with $\psi^\prime\in\mathscr F$, in the sense that
\begin{equation}\label{1.8a}
    \dot \beta_{\theta_0}(\psi,\phi)\geq \dot \beta_{\theta_0}(\psi^\prime,\phi^\prime)
\end{equation}
if
\begin{equation}\label{1.9a}
    \alpha(\psi^\prime,\phi^\prime)\leq \alpha(\psi,\phi)\quad\mbox{and}\quad \mathscr N_{\theta_0}(\psi^\prime)\leq \mathscr N_{\theta_0}(\psi).
\end{equation}
The inequality in (\ref{1.8a}) is strict, if at least one of the inequalities in (\ref{1.9a}) is strict.
If there are equalities in (\ref{1.8a}) and (\ref{1.9a}), then $\psi^\prime$ satisfies (\ref{4.15a})   $\mu^n$-almost everywhere on $T_n^{\psi^\prime}\cap\{f_{\theta_0}^n>0\}$ for all $n=1,2,\dots$ (with $\psi_n^\prime$ instead of $\psi_n$), and $\phi^\prime$ satisfies (\ref{1.7a}) (with $\phi_n^\prime$ instead of $\phi_n$) $\mu^n$-almost everywhere on $S_n^{\psi^\prime}\cap\{f_{\theta_0}^n>0\}$ for all $n=1,2,\dots$.
\end{theorem}

In the same way as in the previous section we can represent the inequalities $g(b-z_n)< c+r_n(b-z_n;c)$ defining the form of the optimal test in a simpler form. Indeed, from Lemma \ref{l11} it is not difficult to deduce that if $c+r_n(z;c)\leq 0$, then in each region $\{z\leq 0\}$ and $\{z\geq 0\}$ there exists a unique solution to the equality
\begin{equation}\label{4.19}
    c+r_n(z;c)=g(z),
\end{equation}
$A_n=A_n(c)\leq 0$   and $B_n=B_n(c)\geq 0$ (see the proof of Lemma \ref{l8}). Let us denote in this case $\Delta_n=\Delta_n(c)=(A_n(c),B_n(c))$  and $\bar \Delta_n=\bar\Delta_n(c)=[A_n(c),B_n(c)]$. In case $c+r_n(z;c)> 0$ let $\Delta_n(c)=\bar\Delta_n(c)=\emptyset$. Then it is easy to see that (\ref{4.15a}) is equivalent to
 \begin{equation}\label{4.20}
    I_{\{b-z_n\in\Delta_n(c)\}}\leq 1-\psi_{n}\leq I_{\{b-z_n\in\bar\Delta_n(c)
\}}.
 \end{equation}
In this way we get the following corollary from  Theorem \ref{t4}.
\begin{corollary}\label{c3}
Under Assumptions  1--3 the assertion of Theorem \ref{t4} remains valid after substituting all the references to  (\ref{4.15a}) for the references to (\ref{4.20}).
\end{corollary}

\begin{remark}\label{r4}\rm
 If in (\ref{4.15a}) (or (\ref{4.20})) and, respectively, in (\ref{1.7a}) $b<0$, then under the conditions of Theorem \ref{t4} (with ``$b<0$'' instead of ``$b>0$'') it follows from Theorem 5.3 \cite{novikov09c} that the test $(\psi,\bar\phi)$, where $\bar\phi_n=1-\phi_n$, $n=1,2,\dots$, is locally most powerful for testing $H_0:\;\theta=\theta_0$ against $H_1:\;\theta<\theta_0$ in the class of all the tests $(\psi^\prime,\phi^\prime)$ for which
$$
E_{\theta_0}\tau_\psi^\prime\leq E_{\theta_0}\tau_\psi\quad\mbox{and}\quad \alpha(\psi^\prime,\phi^\prime)\leq\alpha(\psi,\bar\phi).
$$
 If $b=0$ in (\ref{4.15a}) (or (\ref{4.20})) and in (\ref{1.7a}), then (supposing that all other conditions of Theorem \ref{t4} are fulfilled) the test $(\psi,\phi)$ is locally most powerful for testing $H_0:\;\theta=\theta_0$ against $H_1:\;\theta>\theta_0$, and the test $(\psi,\bar\phi)$ is locally most powerful for testing $H_0$ against  $H_1:\;\theta<\theta_0$, in the class of all tests $(\psi^\prime,\phi^\prime)$ for which
$$E_{\theta_0}\tau_\psi^\prime\leq E_{\theta_0}\tau_\psi
$$
(irrespective of their type-I error probability levels).
\end{remark}

{\bf Some particular cases.} In this section we consider problems of construc-tion of locally most powerful tests in two particular cases  of the general model considered above: in the case  of ``periodic'' process  (see \cite{liu}), and in the case of ``finitely non-stationary'' process of observations (see\cite{novikov08}). The case of i.i.d. observations is a particular case of both of these models.

Let us consider first the    ``periodic'' case, when there exists such natural $T$ that
$f_{\theta,n+T}=f_{\theta,n}$ for all  $n=1,2,\dots$.
In this case, obviously, Assumption 3
is implied by Assumption 1 and 2  (because Assumption  2 guarantees that all
$E_{\theta_0}|\frac{\dot f_{\theta_0,j}}{f_{\theta_0,j}}|$, $j=1,2,\dots, T$, are finite).
It is not difficult to see that
$v_n=v_{n+T}$ and $r_n=r_{n+T}$ for all  $n=1,2,\dots$, so the solutions of the equation  (\ref{4.19}) are also periodical:
$A_n(c)=A_{n+T}(c)$, $B_n(c)=B_{n+T}(c)$, $n=1,2,\dots$.
In addition,
$$
v_{n-1}(z)=\min\{g(z),c+E_{\theta_0}v_n(z-q_n)\}
$$
for all  $n=T,T-1,\dots,2$, and
$$
v_{T}(z)=\min\{g(z),c+E_{\theta_0}v_1(z-q_1)\}.
$$
It is easy to see that in this case
 the sufficient condition of optimality
in Theorem \ref{t4} ($\psi\in\mathscr F$) is also fulfilled, if, additionally to Assumptions 1 -- 2, we assume that
\begin{equation}\label{4.21}
    P_{\theta_0}(\sum_{j=1}^T q_j=0)<1.\end{equation}
Indeed, let  $n=kT$ and $\xi_i=\sum_{j=1}^T q_{(i-1)T+j}$, $i=1,2,\dots$. Then for any  $\psi$, satisfying  (\ref{4.20}), it holds
$$
P_{\theta_0}(\tau_\psi> n)\leq E_{\theta_0}\prod_{j=1}^n I_{\{\sum_{i=1}^j q_i\in b-\bar\Delta_j(c)\}}
$$$$=P_{\theta_0}(\sum_{i=1}^j q_i\in b-\bar\Delta_j(c),\; j=1,2,\dots,n )$$
\begin{equation}\label{4.22}
    \leq P_{\theta_0}(\sum_{i=1}^j \xi_i\in b-\bar\Delta_T(c),\; j=1,2,\dots,k ).
\end{equation}

  Since $\xi_i$, $i=1,2,\dots$  are i.i.d. random variables such that $P_{\theta_0}(\xi_i=0)<1$, the theorem of Stein \cite{stein} applies, due to which, in particular, the right-hand side of  (\ref{4.22}) has an exponential rate of vanishing, as $k\to\infty$. Therefore, $$E_{\theta_0}\tau_\psi=\sum_{n=1}^\infty P_{\theta_0}(\tau_\psi\geq n)<\infty,$$
i.e. $\psi\in\mathscr F$.

If  (\ref{4.21}) is not satisfied, i.e. $P_{\theta_0}(\sum_{j=1}^Tq_j=0)=1$, then, due to independence of $q_j$, $j=1,2,\dots$ we have that $P_{\theta_0}(q_j=0)=1$, for all $j$. By construction, $v_n^N(z)\equiv g(z)$, $r_n^N(z)\equiv g(z)$ for all $N\geq 1$ and for all $n\leq N$, so $v_n(z)\equiv g(z)$, $r_n(z)\equiv g(z)$ for all $n=1,2,\dots$, thus $P_{\theta_0}(\psi_1= 1)=1$ for every $\psi$ satisfying (\ref{4.20}). Therefore, if $(\ref{4.21})$ is not satisfied, then $P_{\theta_0}(\tau_\psi=1)=1$, and $\psi\in\mathscr F$ in a trivial way.

Thus, in the periodic case under Assumptions 1 -- 2 every $(\psi,\phi)$ satisfying (\ref{4.20}) and (\ref{1.7a}) is locally most powerful in the sense of Theorem \ref{t4}.

Let us consider now the ``finitely non-stationary'' case. Let us suppose that  there exists a natural $k$  such that $f_{\theta,j}=f_{\theta,j+1}$, for all  $j\geq k$ ($k=1$ corresponds to the i.i.d. case). Then it is easy to see that  $v_n(z;c)=v(z;c)$, $r_n(z;c)=r(z;c)$ (do not depend on $n$) for all $n\geq k-1$, and, in addition,
\begin{equation}\label{4.21a}
    v(z;c)=\min\{g(z),c+E_{\theta_0}v(z-q_k;c)\},\quad r(z;c)=E_{\theta_0}v(z-q_k;c),
\end{equation}
so the equation  (\ref{4.19}) for determining $A_n(c)$, $B_n(c)$ takes the form:
\begin{equation}\label{4.22a}
      c+r(z;c)=g(z),
\end{equation}
if  $n\geq k-1$. Therefore, $A_n(c)=A(c)$ , $B_n(c)=B(c)$ (do not depend on $n$), if $n\geq k-1$.
For the rest of  $n$ (if any) the recurrent formulas apply:
$$
v_{n-1}(z;c)=\min\{g(z),c+E_{\theta_0}v_n(z-q_n;c)\}, \quad r_{n-1}=E_{\theta_0}v_n(z-q_n;c),
$$
$n=k-1,\dots,1$.

Naturally,  under Assumptions  1 -- 2, and, additionally, the condition
\begin{equation}\label{4.23}
    P_{\theta_0}(q_k=0)<1,
\end{equation}
the  same argument os Stein yields the finiteness of $E_{\theta_0}\tau_\psi$ for every $\psi$ sattisfying (\ref{4.20}). If the conition  (\ref{4.23}) is not fulfilled  (that is, $P_{\theta_0}(q_k=0)=1$), then it follows from (\ref{4.21a}) that  $v(z;c)\equiv g(z)$ and $r(z;c)\equiv g(z)$, so the equation  (\ref{4.22a}) can not have a solution. Thus,  $\Delta_n(c)=\bar\Delta_n(c)=\emptyset$ for all  $n\geq k-1$, which implies that the stopping rule $\psi$ is truncated
($P_{\theta_0}(\tau_\psi\leq k-1)=1$), i.e.  $\psi\in\mathscr F$.
In this way, in the finitely non-stationary case under Assumptions 1 -- 2  every $(\psi,\phi)$ satisfying (\ref{4.20}) and (\ref{1.7a}) is locally most powerful in the sense of Theorem \ref{t4}.

From the  considerations above it is clear that the case $k=2$ is of a special interest  because in this case the boundaries of the continuation region are constant ($A_n(c)=A(c)$, $B_n(c)=B(c)$, $n=1,2,\dots$), so the optimal test has exactly the same structure as in the case of i.i.d observations (see \cite{berk}).
Similar to \cite{berk}, it can be shown in this case (supposing (\ref{4.23}) and the finiteness of the Fisher information $E_{\theta_0}q_2^2$) that for each pair $A<B$, the test $(\psi,\phi)$ for which
 \begin{equation}\label{4.24}
    I_{\{z_n\in (A,B)\}}\leq 1-\psi_{n}\leq I_{\{z_n\in[A,B]
\}}, \; n=1,2,\dots,
 \end{equation}
is locally most powerful.
More precisely, it can be shown that there exist constants $b,\, c$, $A< b< B$, $c>0$ such that (\ref{4.24}) is equivalent to
 \begin{equation}\label{4.25}
    I_{\{b-z_n\in (A(c),B(c))\}}\leq 1-\psi_{n}\leq I_{\{b-z_n\in[A(c),B(c)]
\}},\; n=1,2,\dots,
 \end{equation}
 where $A(c)$, $B(c)$ are solutions of the equation (\ref{4.22a}).

 If the constant $b$ found in this way is positive, $b>0$, then the test $(\psi,\phi)$ with any $\phi$ satisfying (\ref{1.7a}),
 is locally most powerful for testing $H_0:\,\theta=\theta_0$ against $H_1:\,\theta>\theta_0$; if $b<0$, then the test $(\psi,\phi)$ with any $\phi$, satisfying
 $$
I_{\{z_n<b\}}\leq\phi_n\leq I_{\{z_n\leq b\}},\; n=1,2,\dots,
 $$
 is locally most powerful for testing $H_0$ vs. $H_1:\,\theta<\theta_0$; at last, if $b=0$, then both of them are locally most powerful, each for the corresponding alternative (see Remark \ref{r4}).

 As a concluding remark, let us note that if the distribution of $q_2$ is symmetric (as, for example, in the case of normal distribution), then $A(c)=-B(c)$ (see Remark 5.3 in \cite{novikov09c}), so in this case $b=(A+B)/2$.
\vspace{5mm}\\

{\bf  Acknowledgements}. We are very thankful to Alexander Galkin
for providing us with relevant information.

A. Novikov thanks the National System of Investigators (SNI CONACyT), Mexico, for partial support for this work, and also  CONACyT, Mexico,  for partial support under Grant
CB-2005-C01-49854-F.

\end{document}